\documentstyle[preprint,aps]{revtex}
\begin{document}
\title{A non group theoretic proof  of  completeness  of  arbitrary  
coherent  states $D(\alpha)\mid f>$}
\author{G.S. Agarwal}
\address{Physical Research Laboratory\\
Navrangpura\\
Ahmedabad - 580 009 (INDIA)}
\author{S. Chaturvedi}
\address{School of Physics\\
University of Hyderabad\\
Hyderabad - 500 046 (INDIA)}
\maketitle
\begin{abstract}
A new  proof  for  the  completeness  of  the   coherent   states 
$D(\alpha )\mid f>$ for the Heisenberg Weyl group and the  groups  $SU(2)$ 
and $SU(1,1)$ is presented. Generalizations of these results and  their 
consequences are disussed.
\end{abstract}

\newpage
\noindent{\bf Introduction}

Resolution of the identity operator in terms  of  the  eigenstates  of 
suitable operators proves to be an  important  calculational  tool  in 
quantum mechanics. One comes across numerous instances  where  quantum 
mechanical calculations are greatly simplified by a judicious  use  of 
the resolution  of  the  identity  in  terms  of  the  eigenstates  of 
appropriate operators. Among the various resolutions of the  identity, 
the one which has played a key role in quantum optics is that in terms 
of the coherent  states  $\mid\alpha>$  [1-3],  the  eigenstates  of  the 
annihilation operator
\def\bi{{\bf I}}
\def\ad{{a^\dagger}}
\begin{equation}
{1\over\pi} \int d^2\alpha \mid\alpha><\alpha\mid = \bi\,\,\,\,,
\end{equation}
where
\begin{equation}
\mid\alpha> = D(\alpha)\mid0>~~;~~D(\alpha)=\exp(\alpha\ad-\alpha^*a)~~;~~
[a,\ad] = \bi\,\,\,.
\end{equation}

The coherent states $\mid\alpha>$ together with (1) have not only  led  to 
new  calculational  techniques  but  also  led   to   new   conceptual 
developments such as the notion of quasi probability distributions.

The proof of (1) found in  most  text  books  on  quantum  optics  and 
quantum mechanics proceeds by expanding $\mid\alpha>$  in  terms  of  Fock 
states and carrying out the  $\alpha$-integration  and  by  using  the 
completeness  of  Fock   states.   In   recent   times   states   like 
$D(\alpha )\mid n>$, the displaced number states [4-6], have been  used  in 
quantum optics and it is known that these also form a complete set for 
each $n$ [5]. In fact, from a group theoretic point of view [7,8]  one 
has a more general result
\begin{equation}
{1\over\pi}\int    d^2\alpha    D(\alpha)\mid f><f\mid D^\dagger(\alpha)     = 
\bi\,\,\,\,\,,
\end{equation}
where  $\mid f>$,  referred  to  as  the  fiducial  state,  is  any  fixed 
normalizable state. (In  (3)  it  has  been  assumed  that  $\mid f>$  is 
normalized to unity.) The states 
\begin{equation}
\mid \alpha; f> = D(\alpha)\mid f>\,\,\,\,,
\end{equation}
are referred to as generalized coherent states.  (To  avoid  confusion 
with other notions of generlized coherent states, we would, hereafter, 
refer to them as $f$-coherent states.) The choice  $\mid f>=\mid n>$  in  (3), 
for instance, leads to the resolution of the identity in terms of  the 
displaced number states. The group theoretical  proof  of  (3),  using 
Schur's Lemma, is based on the following observations
\begin{itemize}
\item[(a)] $D(\beta)$ provide an irreducible  representation  (upto  a 
phase) of the Heisenberg Weyl group.
\item[(b)] the operator
\begin{equation}
X_1(f)  \equiv  {1\over\pi}\int  d^2\alpha  D(\alpha)\mid f>  <f\mid D^\dagger 
(\alpha) \,\,\,\,,
\end{equation}
commutes with  the  $D(\beta)$'s  and  hence,  by  Schur's  Lemma,  is 
proportional to the identity operator
\begin{equation}
X_1(f) = c(f) \bi\,\,\,\,\,,
\end{equation}
\item[(c)] the constant $c(f)$ can be calculated by taking the  matrix 
element of $X_{1}(f)$  between  any  normalizable  state.  (For  consistency, 
$c(f)$ should be $<\infty$ which,  for  coherent  states  for  certain 
groups  leads  to  restrictions  on  the  fiducial  states.)  For  the 
Heisenberg-Weyl group, it is easy to show that for any fiducial  state 
$\mid f>$; $<f\mid f>=1$, $c(f)=1$ and hence one has (3). By  expanding  $\mid f>$ 
in terms of Fock states (3) may equivalently be written as
\begin{equation}
{1\over\pi}\int d^2\alpha\mid \alpha; n><\alpha;m\mid =\bi  \delta_{nm}~~~;~~~ 
\mid \alpha;n> \equiv D(\alpha)\mid n>\,\,\,\,.
\end{equation}
\end{itemize}

The considerations given above apply to other groups like $SU(2)$  and 
$SU(1,1)$ as well [7,8]. For the case of $SU(2)$
\begin{equation}
[S_+, S_-] = 2S_z~~~ ; ~~~[S_z, S_\pm] = \pm S_\pm \,\,\,\,,
\end{equation}
one has
\begin{equation}
X_2(m) \equiv {2S+1\over4\pi} \int {d^2\zeta\over(1+\mid \zeta\mid ^2)^2} 
\,\,\mid \zeta;m><\zeta;m\mid = \bi\,\,\,\,,
\end{equation}
where
\begin{equation}
\mid \zeta;m> \equiv D(\xi)\mid S,m>~~;  ~~  D(\xi)  =  \exp(\xi  S_+-\xi^*S_-) 
\,\,\,\,,
\end{equation}
and $\mid S,m>$ are eigenstates of $S^2$ and $S_z$. The variables  $\zeta$ 
and $\xi$ are related to each other as follows
\begin{equation}
\xi = {\theta\over2} e^{-i\phi} ~~~ ; ~~~ \zeta  =  \tan{\theta\over2} 
e^{-i\phi} \,\,\,\,\,,
\end{equation}
and the integration in (9) is over the entire $\zeta$-plane.

Similarly, for $SU(1,1)$ 
\begin{equation}
[K_-, K_+] = 2K_z ~~ ; ~~ [K_z, K_\pm] = \pm K_\pm ~~~ ,
\end{equation}
realized via
\begin{equation}
K_+  =  {1\over2}  a^{\dagger2}  ~~  ;  ~~K_-={1\over2}a^2  ~~  ;   ~~ 
K_z={1\over2}(\ad a + {1\over2}) \,\,\,\,,
\end{equation}
one has
\begin{equation}
X_3(n)   \equiv   {1\over2\pi}   \int   {d^2\zeta\over(1-\mid \zeta\mid ^2)^2} 
\mid \zeta;2n+1><\zeta;2n+1\mid = \bi_{odd} \,\,\,\,,
\end{equation}
where
\begin{equation}
\mid \zeta;2n+1>\equiv   D(\xi)\mid 2n+1>~;~D(\xi)=\exp(\xi   K_+-\xi^*K_-)~;~ 
K_z\mid 2n+1>=(n+{3\over4})\mid 2n+1> \,\,\,\,,
\end{equation}
and $\zeta$ and $\xi$ are related to each other as follows
\begin{equation}
\xi=\mid \xi\mid e^{-i\phi}~~~ ;~~~ \zeta=\tanh\mid \xi\mid e^{-i\phi}~~~.
\end{equation}
The operator $\bi_{odd}$ in (14) denotes the unit operator in the  odd 
sector of the Fock space.
\begin{equation}
\bi_{odd} \equiv \sum_{k=0}^\infty \mid 2k+1><2k+1\mid \,\,\,\,,
\end{equation}
and the integration in (14) is over the  unit  disc  centered  at  the 
origin in the complex $\zeta$-plane.

\noindent{\bf New proof of completeness of $f$-coherent states}

We first consider (3). To prove (3) in a rather elegant way we  make 
use of the following results:
\begin{itemize}
\item[(i)] resolution of the identity (1) in terms of coherent states.
\item[(ii)] the fact that an operator is  uniquely  determined  by 
its diagonal elements [9].
\end{itemize}
\begin{equation}
<\beta\mid G\mid \beta> = 1 ~\mbox{for all} ~ \beta ~~\mbox{if and only if} ~~
G = \bi\,\,\,\,.
\end{equation}
Now consider the operator $X_1(f)$
\begin{equation}
X_1(f)    \equiv    {1\over\pi}    \int    d^2\alpha     D(\alpha) 
\mid f><f\mid D^\dagger(\alpha)\,\,\,\,.
\end{equation}
Consider the diagonal elements of $X_1(f)$
\begin{eqnarray}
<\beta\mid X_1(f)\mid \beta>   &    =    &    {1\over\pi}\int    d^2\alpha 
<\beta\mid D(\alpha)\mid f><f\mid D^\dagger(\alpha)\mid \beta>\,\,\,\,,\nonumber\\
& = & 
{1\over\pi}\int    d^2\alpha 
<0\mid D^\dagger(\beta)D(\alpha)\mid f><f\mid D^\dagger(\alpha)D(\beta)\mid 0>\,\,\,\,,
\end{eqnarray}
which on using the algebraic property of the displacement operator 
$D(\alpha)$
\begin{equation}
D^\dagger(\beta)D(\alpha) =  D(\alpha-\beta) \exp[(\beta^*\alpha - 
\beta\alpha^*)/2]\,\,\,\,,
\end{equation}
reduces to
\begin{equation}
<\beta\mid X_1(f)\mid \beta>         =          {1\over\pi}          \int 
d^2\alpha\mid <0\mid D^\dagger(\beta-\alpha)\mid f>\mid ^2\,\,\,\,.
\end{equation}
On rewriting the integrand (22) in terms of  coherent  states  and 
changing the variable of integration (22) becomes
\begin{eqnarray}
<\beta\mid X_1(f)\mid \beta>     &     =      &      {1\over\pi}      \int 
d^2\alpha\mid <\beta-\alpha\mid f>\mid ^2\nonumber\\
& = & {1\over\pi} \int d^2\alpha<f\mid \alpha><\alpha\mid f>\nonumber\\
& = & <f\mid {1\over\pi}\int d^2\alpha\mid \alpha><\alpha\mid f> = 1 \,\,\,\,.
\end{eqnarray}
Thus the diagonal coherent elements of $X_1(f)$ for all values  of 
$\beta$ are equal to unity and therefore using the  property  (18) 
we conclude that
\begin{equation}
X_1(f) = \bi\,\,\,\,.
\end{equation}
This constitutes  a  direct  proof  of  the  completeness  of  the 
$f$-coherent states of the Heisenberg-Weyl group.

Next we consider the $SU(2)$ case. In this the  analogues  of  (i) 
and (ii) above are

\noindent  (i)  ~~completeness  of  the  atomic  coherent   states 
$\mid \zeta;-S>$ [10]
\begin{equation}
{2S+1\over4\pi}                 \int{d^2\zeta\over(1+\mid \zeta\mid ^2)^2} 
\,\,\mid \zeta;-S><\zeta;-S\mid = \bi \,\,\,\,,
\end{equation}
(ii) \hfill $<\zeta;-S\mid G\mid \zeta;-S> = 1$ for all $\zeta$ if an only 
if $G=\bi$. \hfill (26)

\noindent We consider the diagonal  matrix  elements  of  $X_2(m)$ 
defined   in   (9)   between   the    atomic    coherent    states 
$\mid \zeta^\prime;-S>$. We follow the same procedure as above and use 
the following algebraic properties.
\addtocounter{equation}{1}
\begin{equation}
D(\xi_1)D(\xi_2) =  D(\xi_3)\exp[i\Phi(\xi_1,\xi_2)S_z]\,\,\,\,,
\end{equation}
where
\begin{equation}
\Phi(\xi_1,\xi_2) = {1\over  i}  \ln\left[{1-\zeta_1\zeta_2^*\over 
1-\zeta_1^*\zeta_2}\right]\,\,\,\,,
\end{equation}
and
\begin{equation}
\zeta_3 = {\zeta_1+\zeta_2\over1-\zeta_1^*\zeta_2} \,\,\,\,.
\end{equation}
Further, under the change of variables from $\zeta_2$ to $\zeta_3$ 
the measure of integration in (9) is invariant
\begin{equation}
{d^2\zeta_2\over(1+\mid \zeta_2\mid ^2)^2}                               = 
{d^2\zeta_3\over(1+\mid \zeta_3\mid ^2)^2} \,\,\,\,.
\end{equation}
Using these relations we obtain
\begin{equation}
<\zeta^\prime;-S\mid X_2\mid \zeta^\prime;-S>={2S+1\over4\pi}          \int 
{d^2\zeta^{\prime\prime}\over(1+\mid \zeta^{\prime\prime}\mid ^2)^2} 
<S,N\mid \zeta^{\prime\prime};-S><\zeta^{\prime\prime};-S\mid S,N> \,\,\,,
\end{equation}
which, on using the completeness of  the  atomic  coherent  states 
yields
\begin{equation}
<\zeta^\prime;-S\mid X_2(m)\mid \zeta^\prime;-S>   =    1
~~\mbox{for    all}~~
\zeta^\prime \,\,\,\,,
\end{equation}
and hence $X_2(m)=\bi$. It is important to note that the  fiducial 
state in this case must be an eigenstate of  $S_z$  otherwise  the 
phase factor which arises from the use of (27) will not cancel.

Similarly, in the $SU(1,1)$ case, we use the  following  algebraic 
properties.
\begin{equation}
D(\xi_1)D(\xi_2) = D(\xi_3)\exp[i\Phi(\xi_1,\xi_2)K_z]\,\,\,\,, 
\end{equation}
where
\begin{eqnarray}
\Phi(\xi_1,\xi_2) &  = &  {1\over  i}  \ln\left[{1+\zeta_1\zeta_2\over 
1+\zeta_1^*\zeta_2}\right] \,\,\,\,,\\
\zeta_3 & = & {\zeta_1+\zeta_2\over1+\zeta_1^*\zeta_2} \,\,\,\,.
\end{eqnarray}
The measure of  integration  is  invariant  under  the  change  of 
variables from $\zeta_2$ to $\zeta_3$
\begin{equation}
{d^2\zeta_2\over(1-\mid \zeta_2\mid ^2)^2}                               = 
{d^2\zeta_3\over(1-\mid \zeta_3\mid ^2)^2} \,\,\,\,.
\end{equation}
On using the completeness of $\mid \zeta;1>$, one can show that
\begin{equation}
<\zeta;1\mid X_3(n)\mid \zeta;1> = 1~~ \mbox{for all} ~~\zeta\,\,\,\,,
\end{equation}
and hence $X_3(n)=\bi$.

\noindent{\bf Outlook:}

We have thus shown that
\begin{equation}
\int d\mu(\zeta) D(\zeta)\mid f><f\mid D^\dagger(\zeta) = \bi \,\,\,,
\end{equation}
for the $f$-coherent states for the three groups considered above. 
The relation (38) is amenable to further generalisations. In the 
case of Heisenberg- Weyl group, by expanding the state $\mid f>$ in 
(38) in terms of the number states $\mid n>$ one obtains
\begin{equation}
\int   d\mu(\zeta)   D(\zeta)\mid m><n\mid D^\dagger(\zeta)   =    \bi 
\delta_{mn} \,\,\,\,,
\end{equation}
and hence
\begin{equation}
\int   d\mu(\zeta)   D(\zeta)\mid f_1><f_2\mid D^\dagger(\zeta)   =    \bi 
<f_1\mid f_2> \,\,\,\,.
\end{equation}
In view of (39), one has
\begin{equation}
\int d\mu(\zeta) D(\zeta)\rho_o D^\dagger(\zeta) = \bi \,\,\,\,,
\end{equation}
where $\rho_o$ is an arbitrary density matrix.   
For $SU(2)$ and $SU(1,1)$, (38) implies (41) with $\rho_o$ subject to
the conditions
\begin{equation}
[\rho_o, S_z] = 0 ~~\mbox{and} ~~[\rho_o, K_z] = 0\,\,\,\,,
\end{equation}
respectively. It may be noted that, in the context of Heisenberg-Weyl group,
resolutions of the identity of the type (41) have been derived by Vourdas and 
Bishop [11] for two specific choices of $\rho_o$. The fact that, for the
Heisenberg-Weyl group (41) is valid for an arbitrary $\rho_o$ does not seem 
to be generally appreciated.
 
The results given above enable us to derive  interesting identities 
involving  orthogonal  polynomials. For  example  the   following 
integral\footnote{A direct proof of (43) appears to be  difficult. 
We  have  succeeded  in  proving   it   using   Racah   identities 
[13].} involving 
the     Jacobi polynomials $P_n^{(\alpha,\beta)}(x)$ [12]
\begin{equation}
{1\over2}    \left[{\Gamma(n+1)~    \Gamma(p+3/2)\over\Gamma(p+1)~ 
\Gamma(n+3/2)}\right] ~ \int_o^1  {dx\over(1-x)^{1/2}}  ~  x^{p-n} 
\left[ P_n^{p-n,1/2)} (1-2x)\right]^2 = 1 \,\,\,\,,
\end{equation}
can be derived from (38) by applying it to the $SU(1,1)$ case and using the 
relations\footnote{Expressions for these matrix elements in  terms 
of associated Legendre functions may be found in [7].}
\begin{eqnarray}
<2m+1\mid D(\xi)\mid 2n+1> & =  &  e^{-i(m-n)\phi}  \left[{\Gamma(n+1)   
\Gamma(m+3/2)\over\Gamma(m+1)    ~     \Gamma(n+3/2)}\right]^{1/2} 
(\mid \zeta\mid )^{m-n}(1-\mid \zeta\mid ^2)^{3/4}\nonumber\\
\nonumber\\
&&~~~~~~~~~~~~~~~~ P_n^{(m-n,1/2)}(1-2\mid \zeta\mid ^2)  ~~\mbox{for}  ~~ 
m\ge n\,\,\,\,,\\
\nonumber\\
& =  &  e^{-i(n-m)\phi}  \left[{\Gamma(m+1)  ~ 
\Gamma(n+3/2)\over\Gamma(n+1)    ~     \Gamma(m+3/2)}\right]^{1/2} 
(-\mid \zeta\mid )^{n-m}(1-\mid \zeta\mid ^2)^{3/4}\nonumber\\
\nonumber\\
&&~~~~~~~~~~~~~~~~ P_m^{(n-m,1/2)}(1-2\mid \zeta\mid ^2)  ~~\mbox{for}  ~~ 
m\le n\,\,\,\,,
\end{eqnarray}
In conclusion, we also note the possibility of using relations like (1) 
to construct new classes of quasi-probability distributions. Thus, for
instance, for any density operator $\rho$, one can define a generalised 
Q-function as follows
\begin{equation}
Q(\zeta) = Tr[\rho D(\zeta)\rho_o D^\dagger(\zeta)]
\end{equation}
We hope to discuss this in detail elsewhere.

\end{document}